\documentclass[twocolumn]{svjour3}         
\smartqed  
\usepackage{graphicx}
\begin{document}

\title{Studies of Dense Cores with ALMA}


\author{Mario Tafalla}


\institute{Mario Tafalla\at
              Observatorio Astron\'omico Nacional, Alfonso XII 3, E-28014 Madrid, 
	      Spain \\
              \email{m.tafalla@oan.es}           
}

\date{Received: date / Accepted: date}

\maketitle

\begin{abstract}
Dense cores are the simplest star-forming sites
that we know, but despite their simplicity, they
still hold a number of mysteries that limit our 
understanding of how solar-type stars form.
ALMA promises to revolutionize our knowledge of
every stage in the life of a core,
from the pre-stellar phase to the final disruption
by the newly born star.
This contribution presents a brief review
of the evolution of dense cores and illustrates 
particular questions that will greatly benefit 
from the increase in resolution 
and sensitivity expected from ALMA.


\end{abstract}

\section{Introduction}
\label{intro}

Nearby dark clouds like Taurus and Perseus contain dozens
of dense molecular cores where stars like our Sun are
currently forming or have done so in the recent past
(Myers 1995). Their large number, together
with their proximity and simple structure, 
make cores unique targets to study the complex physics
involved in the formation of a star. Dense cores
that have not yet formed stars, the so called
starless or pre-stellar cores, inform us of the
initial conditions of star formation, and their
study can help us elucidate the process by
which pockets of cloud material condense 
and become gravitationally
unstable. Cores with deeply embedded young stellar objects
(``protostellar cores'') are unique
targets to study the complex motions that occur during
the period of accretion, when
a combination of infall, outflow, and rotation 
is necessary to assemble the star and redistribute
the gas angular momentum. Finally, evolved cores
are primary targets to study the interaction
between the newly born star and its environment.
These feedback effects are responsible for the transition
of the protostar from embedded to visible, and may be
important determining the final mass of the star
and stabilizing the nearby gas via turbulence generation.

The observational study of dense cores has advanced enormously over
the last decade thanks to the increase in resolution
provided by the new millimeter and submillimeter 
interferometers, and also due to the systematic
combination of observations of dust and molecular 
tracers (e.g., Bergin \& Tafalla 2007). This brief review summarizes
some new results from dense cores studies
and presents a number of current issues that will
greatly benefit from ALMA observations.
The limited space of this article makes any attempt
to review the field 
necessarily incomplete, and the reader is
referred for 
further information to the other contributions on
star formation in these proceedings,
in particular to those
by van Dishoeck, Andr\'e, Shepherd,
Aikawa, Wilner, Johnstone, and Crutcher.

Despite significant recent progress, our understanding of the
structure and evolution of dense cores is still
incomplete due in part to limitations in the resolution and 
sensitivity of the available observations. Even the highest resolution
data of nearby dense cores cannot discern
details finer than about 100 AU, which is still
insufficient to disentangle the complex kinematics
of infall and outflow motions in the vicinity
of a protostar. Probably more important, the
low temperatures of the gas and the dust in
cores ($\approx 10$~K) 
make the emission of any core tracer intrinsically weak,
so any increase in the resolution needs to be
accompanied by a parallel increase in the 
sensitivity, or the observations will not achieve enough 
S/N to provide useful information. This is particularly
important when using weak, optically thin tracers to sample
the innermost gas in the core. These tracers, in 
addition, often present extended emission,
which poses a problem to the current generation of
interferometers that cover sparsely the $uv$ plane
and therefore suffer systematically from missing
flux. The high resolution and collecting area afforded by ALMA, 
combined with its great sensitivity to extended emission,
promises to revolutionize the field of dense cores studies.
On the one hand, ALMA will allow studying the dense cores 
of nearby clouds with the greatest detail, achieving subarcsecond
resolution with high sensitivity. On the other hand, ALMA will
permit the systematic study of dense cores in more distant clouds, enlarging 
the sample of available targets from the current 
set of the nearest clouds to cores at distances of at least 1~kpc.

\begin{figure*}
\begin{center}
\includegraphics[angle=-90,width=0.75\textwidth]{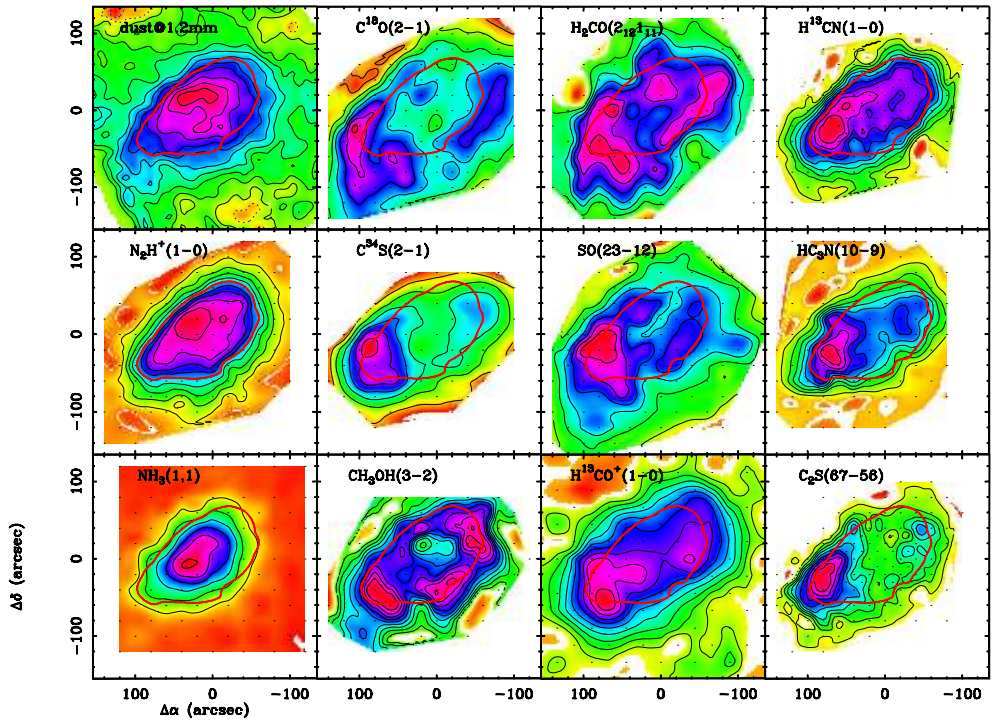}
\caption{Sample of maps of the L1498 dense core in Taurus illustrating
its differentiated chemical composition. The three panels in the left
show tracers that are sensitive to the core interior,
and therefore present a centrally-concentrated emission
(dust continuum map is in the top left panel). All other panels show
ring-like distributions of emission that result from the depletion of
the emitting molecules in the core interior (data from Tafalla et al. 2006).
}
\label{fig:1}
\end{center}
\end{figure*}

\section{Pre-stellar cores}
\label{sec:2}

The earliest phase of a core, the so-called starless or
pre-stellar stage, is characterized by the lack of 
a point-like object at its center (e.g., Di Francesco et al. 2007). This 
characterization is of course dependent 
on the current sensitivity limits of the observations, and is therefore 
susceptible of misclassifying a core with an embedded
source of very low luminosity (see the case
of VeLLOs below). Still, the significant number of dense
cores with no pointlike source detected even after 
deep Spitzer Space Telescope observations suggests 
that a population of truly starless cores exists in nearby
clouds like Taurus (Werner et al. 2006). 

Starless cores present systematically a close to
constant density of $10^5$-$10^6$ cm$^{-3}$ over the 
central 5000-10000 AU followed by an almost
power-law drop at large distances.
This central flattening of the density profile has been 
observed in a number of cores using different
observational techniques, like millimeter dust continuum 
emission (Ward-Thompson et al. 1999), 
MIR absorption (Bacmann et al. 2000), and
NIR extinction (Alves et al. 2001), and therefore
constitutes a robust result of recent core studies.
The presence of a density flattening provides further evidence 
that starless cores 
have not yet developed a central singularity,
and that they are of pre-stellar nature. The
physical origin of the flattening, however, is still
a matter of debate, as a number of interpretations 
are consistent with it. The most natural one
is that the profile results from an equilibrium
configuration in which the pressure of an
isothermal gas balances its gravitational attraction,
the so called Bonnor-Ebert profile (e.g.,
Alves et al. 2001). Indeed, the gas temperature
in a core is typically close to constant
($\approx 10$~K),
and the associated thermal pressure dominates
the turbulent component by a factor of several
(e.g., Tafalla et al. 2004). The Bonnor-Ebert interpretation,
however, seems in conflict with the non-spherical
shape of most cores (typical axial ratio is 2:1, Myers
et al. 1991), and with the fact that the density contrast 
observed in cores often exceeds the factor of 14
limit for stability of the Bonnor-Ebert analysis
(Bacmann et al. 2000). Additional magnetic
field support could be responsible for these deviations
from the theoretical expectation, 
but unfortunately, the observation of
this magnetic component is extremely hard to make
(see contribution from Crutcher in this volume). 
Even the apparently ``simple'' structure of the 
cores still eludes our understanding.

When the density distribution of a core, as inferred from dust
measurements, is compared with the observed emission from
most molecular tracers, it is commonly found that they
disagree significantly. As illustrated in
Fig.~1 for L1498 in Taurus, the dust emission
of a core often appears centrally concentrated 
(with of course a relative flattening at the center),
while all molecular species but NH$_3$ and N$_2$H$^+$
present ring-like distributions around the continuum
peak. Radiative transfer analysis of the molecular
emission indicates that the abundance of most species
drops by at least a factor of 10 towards the high density
peak of the molecular core (Caselli et al. 1999, 
Bergin et al. 2002, Tafalla et al. 2002).
Such strong abundance decrease is suffered
by all the C-bearing molecules as well as other species
(like SO), while it does not affect significantly 
NH$_3$ or N$_2$H$^+$
(see Di Francesco et al. 2007 and Bergin \& Tafalla 2007 
for reviews).
NH$_3$ seems in fact to be enhanced toward the center
of most cores (Tafalla et al. 2002),
while the N$_2$H$^+$ abundance tends to have a 
constant value or may drop at the very center of
some cores (Bergin et al. 2002, Pagani et al. 2005). 
Cores therefore have
a differentiated (onion-like) molecular composition,
with a center rich in NH$_3$ and N$_2$H$^+$ and 
a series of outer layers containing C-bearing
species.

The inhomogeneous composition of the starless dense cores
most likely results from the freeze out of the main 
molecular species onto the cold dust grains at the 
center (Bergin \& Langer 1997, Aikawa et al. 2005).
The high densities and low temperatures typical
of dense core centers
make the freeze out time ($\approx 5\; 10^9/n_{H_2}$ yr)
become much shorter than the core dynamical scale
($\approx$ 1~Myr), and as a consequence, species like
CO disappear rapidly from the gas phase. 
Other molecular species suffer the same fate as
CO, but more importantly, the original chemical
balance, characterized by a relative large 
CO abundance ($\sim 10^{-4}$), is changed dramatically
by freeze out.
A new chemical balance emerges, and it is characterized by
the enhancement of certain N-bearing species, like N$_2$H$^+$, which
are daughter products of N$_2$ and whose
abundance is controlled by the 
amount of CO in the gas phase (CO is the main 
destroyer of  N$_2$H$^+$). Even as N$_2$ freezes out on the 
dust grains with a similar binding energy as
CO ({\"O}berg et al. 2005), the N$_2$H$^+$ abundance
can increase relatively from its value in the
diffuse cloud (where CO is undepleted) and give rise to
the relatively ``high'' abundances
(few 10$^{-10}$) typical of dense cores. 
NH$_3$ can then form from N$_2$H$^+$ via 
dissociative recombination (Geppert et al. 2004),
giving rise to the observed central enhancement 
(Aikawa et al. 2005).

Another effect of the CO depletion
in cores is the enhancement
of deuterated species. Deuteration 
at the low (10~K) temperature of dense cores
occurs via the enhancement of 
H$_2$D$^+$, which then passes the deuterium
atom to other species via ion-molecule
reactions (Dalgarno \& Lepp 1984).
As H$_2$D$^+$ is mainly destroyed by CO,
the depletion of CO 
further enhances the H$_2$D$^+$ abundance,
which in turn enriches in deuterium a
number of additional species.
High abundance of H$_2$D$^+$ has in fact been 
observed in the heavily CO-depleted dense core 
L1544 (Caselli et al. 2003), and a correlation of
CO depletion and high deuteration has been 
reported by Bacmann et al. (2003) and
Crapsi et al. (2005). This deuteration in the 
cold and dense pre-stellar
phase is responsible for the extreme deuteration
values of species like H$_2$CO, CH$_3$OH, and NH$_3$
seen toward protostellar cores (Ceccarelli et al. 1998,
Roueff et al.  2000, van der Tak et al. 2002)

\section{From cores to protostars}

As cores evolve, they are expected to become more
and more centrally concentrated until they
reach the point of gravitational instability.
One of the most pressing issues in star formation
studies is to understand whether this process
of concentration is driven by the loss of magnetic
field support via ambipolar diffusion (e.g.,
Shu et al. 1987, Mouschovias \& Ciolek 1999)
or by the dissipation of turbulence via
shocks (e.g., MacLow \& Klessen 2004). Observations
of dense cores cannot yet distinguish between
these scenarios, but do show a systematic correlation
between central concentration and other indicators of
evolution, like CO depletion and deuterium fractionation
(Crapsi et al. 2005). Evidence for inward motions
also seems correlated with central concentration, 
and this suggests that some cores that we see now as 
starless have already begun collapsing to
form stars. One of the best candidates for
such a collapsing system is the L1544 core
in Taurus, whose pattern of inward motions
has been studied in a number of molecules
(Tafalla et al. 1998, Williams et al. 1999,
Caselli et al. 2002). The L1544 dense core
is characterized by a high central density and concentration
(Ward-Thompson et al. 1999, Tafalla et al. 2002),
a high degree of CO depletion and deuterium
fractionation (Caselli et al. 1999, 2002),
and seems starless despite deep Spitzer
Space Telescope observations in the IR
(Bourke, private communication). Clearly
this core, an similar objects, will be prime targets
for ALMA observations.

Cores more evolved than L1544 are expected to contain already
a luminous object surrounded by an envelope of accreting
material. The little observable difference between 
the pre and proto-stellar phases of a core is illustrated
by the case of L1521F, a core initially thought from
molecular data to be an almost twin of L1544 (Crapsi et al.
2004) and later found with Spitzer observations to have a 
luminous central star (Bourke et al. 2006). The central object in 
L1521F has a luminosity close to 0.1~L$_\odot$, and
is characteristic of a new group of objects
identified by the Spitzer telescope and
usually referred as VeLLOs (Very Low Luminosity
Objects). These VeLLOs seem associated
with very weak NIR nebulosity and low
velocity bipolar outflows (Bourke et al. 2005),
and their status in the evolutionary sequence of
protostars is still unclear. Although some VeLLOs could
represent precursors of substellar objects
(''proto brown dwarfs''), it seems more likely that
in the case of L1521F we are witnessing the
very first moments of accretion, when the central
source has an extremely low mass. 
The proto brown dwarf alternative is unlikely
in this case because 
the dense core has about 5 M$_\odot$ of mass 
(Crapsi et al. 2004), and no clear perturbation
seems stopping the accretion 
(the outflow has too little mechanical power).

The pristine nature of VeLLOs makes them 
ideal candidates to study 
star-forming infall motions. The study of 
these motions has a long and rich
tradition, and is plagued by difficulties
as illustrated by the case of B335. This
dense core harbors a very young (Class 0)
object whose inward motions were first characterized
by Zhou et al. (1993). These authors found
that the spectral signatures from this core are 
in good agreement with the expectation from the
inside-out collapse model of Shu (1977). High resolution
observations with the Plateau de Bure Interferometer by 
Wilner et al. (2000), however, have shown
that some of the signatures of ``infall'' (like the high velocity
wings in the CS lines) arise in fact from outflow 
acceleration, and not from an increase in velocity
of the infalling material as it approaches the central
object. A revisit of B335 (and similar objects)
making use of ALMA's  high angular resolution and selecting 
appropriate (i.e., depletion resistant) tracers
is therefore needed to clarify the still confusing picture
of star-forming infall motions. The clean
appearance of some VeLLOs, together with their
weaker outflow emission, offers an interesting 
alternative to the more evolved (and massive) objects
like B335, that have fully developed outflows. Because 
of their lower mass, VeLLOs
may present weaker signatures of infall and may be tracing
the very first moments of collapse. The combined study of 
VeLLOs and more luminous Class 0 and Class I sources
should therefore allow us to reconstruct the sequence of star-forming
accretion as a function of time.

The presence of a protostar at the center of a
core affects not only the gas kinematics
but its chemistry. The newly born star
heats up the nearby gas and dust 
introducing a temperature gradient in its vicinity.
In the $\sim 1000$~AU region where the dust temperature 
exceeds the CO evaporation temperature
($\approx 20$-30~K), this molecule
returns to the gas phase and undoes 
part of the chemical processing that occurred during the 
pre-stellar phase (J{\o}rgensen et al. 2004,
J{\o}rgensen 2004).
Closer to the protostar ($\sim 100$~AU), the dust temperature 
reaches the 90-100~K value at which water  
evaporates from the grains, further enriching 
the chemistry.
Observations of some very young protostellar objects, like IRAS 
16293-2422, show that these very small regions have
extreme abundance of a number of complex molecules
like HCOOH, HCOOCH$_3$, and CH$_3$OCH$_3$
(Cazaux et al. 2003, Bottinelli et al. 2004).
The chemical richness of these regions rivals
that of the hot cores around massive protostars,
justifying their common denomination as 
``hot corinos'' (Ceccarelli et al. 2007). The exact
origin of the complex molecules in these regions,
however, is still not fully understood. One possibility
is that they result from direct evaporation of 
species trapped in the water ice, while an alternative
is that they result from the processing of
simpler evaporated molecules. Even the geometry of
hot corinos remains unknown, with the innermost part of the
envelope or a more stable disk-like distribution
as the most likely locations.
Despite these temporary uncertainties, hot corinos
offer a unique opportunity to study the innermost
vicinity of low-mass protostars. Their distinctive
chemical composition makes them highly selective
tracers of the most complex and interesting region
of the protostar, where inflow, outflow, and rotation
motions play comparable roles, and angular momentum
is transfered between different gas components. Hot corino
studies with ALMA will surely constitute some of
the first scientific projects of the instrument.

\section{Outflow acceleration and core disruption}

\begin{figure*}
\begin{center}
\includegraphics[width=0.75\textwidth]{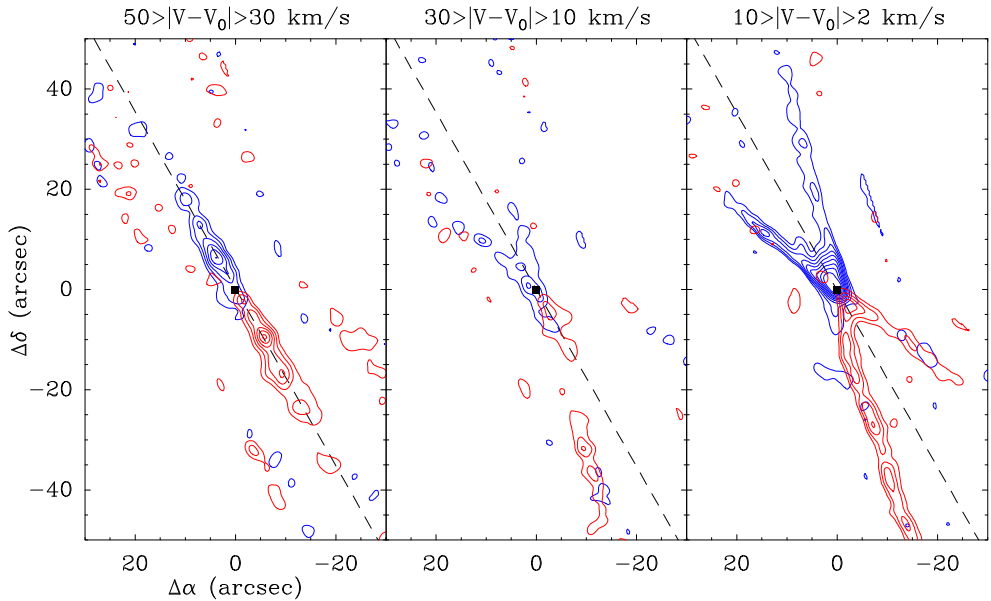}
\caption{CO(J=2--1) emission from the IRAS 04166+2706 outflow
(PdBI data, Santiago-Garc\'{\i}a et al. in preparation). The highest velocity
gas (left panel) forms two opposed jets that emerge from the IRAS source
(filled square)
and travel toward the north-east (blue gas) and south-west (red gas).
The intermediate velocity regime (middle panel) is almost absent, while
the low velocity gas (right panel) lies along the walls of
two opposed evacuated cavities. The combination of highly collimated
jets and limb-brightened shells requires an underlying wind
with both highly-collimated and wide-angle components, and shows the
limitation of single-component models. See poster contribution from
Santiago-Garc\'{\i}a et al. for further details.
}
\label{fig:2}
\end{center}
\end{figure*}

At the same time that protostars accrete material, 
they eject powerful bipolar outflows of supersonic speed.
CO observations of these outflows reveal masses that are too large 
to originate directly from the central protostar, and indicate that
most of the moving gas is core ambient material accelerated by
a collimated stellar wind (Lada 1985). The lobes of bipolar outflows, in addition, 
commonly coincide with evacuated cavities seen via scattered
light from the protostar, further illustrating how the
outflow phenomenon represents a major disruption in 
the core internal structure (Padgett et al. 1999).

Despite more than two decades of intense outflow research,
a number of outstanding problems remain, and ALMA observations
represent our current best hope to solve them (see also contribution by 
D. Shepherd in this volume). The properties of the
underlying wind, for example, are not yet understood, and
several alternative models have been proposed over the years.
The two main types of models that attempt to fit the observations 
are the jet-driven outflow 
and the wind-driven shell, each of them with a number of
flavors (see  Bachiller 1996 for a review).
Despite significant successes, however, neither type of model can 
reproduce the rich variety of kinematic properties found by 
observations, so each of of them is necessarily incomplete 
(Lee et al. 2002). In the jet driven model,
a highly collimated agent shocks and sweeps cloud material
along an almost straight line. This model succeeds
in explaining the highly collimated CO outflows often found
toward Class 0 objects, but fails to reproduce observations
of less collimated flows (usually powered by Class I sources),
where the CO emission arises from gas along limb-brightened 
shells (like L1551, see Moriarty-Schieven et al. 1987). 
To fit these less collimated systems,
the jet models need to broaden the outflow path, and
this has been done by either invoking jet 
precession/''wandering'' (Masson \& Chernin 1993) or large-scale
bow shocks (Raga \& Cabrit 1993). None of these elements however
seems consistent with observations (see Arce et al. 2007 for
more details), and this leaves the jet models limited to fitting the
youngest, and admittedly more spectacular, bipolar outflows.
Wind-driven models, on the other hand,
naturally produce shell-like structures
thanks to a wide-angle agent that sweeps ambient material
(Shu et al. 1991). These models, unfortunately, do not reproduce
the appearance of the highly collimated outflows or the 
mass-velocity distribution commonly observed even in the poorly collimated 
flows (Masson \& Chernin 1992).

A combination of high resolution observations and new developments
in outflow modeling are starting to show a possible
solution to the current impasse. Interferometer mapping
of the outflow powered by the very young source 
IRAS 04166+2706 in Taurus shows both jet and
shell features simultaneously (see Fig. 2 and poster contribution by 
Santiago-Garc\'{\i}a et al.). The jet-like feature in this outflow,  
seen in both CO and SiO emission, is extremely rectilinear, appears 
only at the highest velocities (between 30 and 50 km s$^{-1}$), and shows 
no evidence for precession or wandering. The shell-like part appears 
at low velocities (2 to 10 km s$^{-1}$) and seems to delineate two opposed
cavities with the IRAS source at their vertex. This cavity interpretation
is supported by the fact that the blue outflow shell
coincides with the walls of a NIR scattering nebula seen in Spitzer images,
as expected from its more favorable projection. In addition, the high velocity jet
runs along the axis of the two cavities showing a remarkable degree of symmetry
(see poster contribution for further details). The data from IRAS 04166+2706,
therefore,
leads to the inevitable conclusion that, at least in some cases, both
highly collimated and wide-angle components coexist in the outflow
driving agent, and that a model that considers both components simultaneously
is needed to explain the observations. Interestingly enough, recent
realistic modeling of the interaction between the X-wind of Shu et al.
(1994) and a toroidal core shows that both jet
and shell components should be observed simultaneously in 
very young outflows (Shang et al. 2006).
This so-called ``unified'' model of bipolar flows shows in fact a remarkable
likeness with the IRAS 04166+2706 observations, both in geometry and
kinematics (compare Fig.~2 and the models in Shang et al. 2006).

The unified outflow model not only unifies the jet and wide-angle
aspects of the outflows, but also brings together the evolution of 
flows and the dense cores, two elements often treated separately.
Evidence for outflow-core interaction has been reported in a number of
systems (e.g., Tafalla \& Myers 1997, Arce \& Sargent 2006), but no unified
framework of how this interaction happens or how outflows
and cores evolve in parallel exists yet. The beautiful simulations of
Shang et al. (2006) illustrate how the most important elements of this 
interaction occur inside the central 1000~AU region, which corresponds
to less than $10''$ even towards the most nearby clouds. High angular
resolution observations with ALMA are clearly needed to sample the
complex geometry and kinematics inside this critical region, and thus 
compare real outflows with their simulated counterparts. Producing a 
unified picture of the different and interacting processes occurring
during the formation of a solar-type star can be one of  
most significant achievements of ALMA.

\end{document}